\documentclass[prd,aps,floatfix,superscriptaddress,twocolumn,nofootinbib]{revtex4-2}

\usepackage{enumerate}
\usepackage{bm}
\usepackage{bbm}
\usepackage[usenames,dvipsnames]{xcolor}
\usepackage{amssymb,amsmath}
\usepackage{graphicx}
\usepackage{overpic}
\usepackage{color}
\usepackage[colorlinks=true,citecolor=magenta]{hyperref}
\usepackage{enumitem}
\usepackage{braket}
\usepackage{tensor}

\def\ba{\begin{eqnarray}}
\def\ea{\end{eqnarray}}
\newcommand{\mpl}{M_{\rm{Pl}}}
\newcommand{\pd}{\partial}

\newcommand{\cut}{\Lambda}
\newcommand{\G}{{\cal G}}
\newcommand{\K}{{\cal K}}

\renewcommand{\epsilon}{\mu}
\def\d{\mathrm{d}}
\def\({\left(}
\def\){\right)}

\makeatletter 
    
\renewcommand\onecolumngrid{
\do@columngrid{one}{\@ne}
\def\set@footnotewidth{\onecolumngrid}
\def\footnoterule{\kern-6pt\hrule width 1.5in\kern6pt}
}

\renewcommand\twocolumngrid{
        \def\footnoterule{
        \dimen@\skip\footins\divide\dimen@\thr@@
        \kern-\dimen@\hrule width.5in\kern\dimen@}
        \do@columngrid{mlt}{\tw@}
}%

\makeatother    

\begin{document}

\title{Causality Constraints on Gravitational Effective Field Theories}

\author{Claudia de Rham}
\email{c.de-rham@imperial.ac.uk}

\author{Andrew J.~Tolley}
\email{a.tolley@imperial.ac.uk}

\affiliation{Theoretical Physics, Blackett Laboratory, Imperial College London, SW7 2AZ London, United Kingdom}

\author{Jun Zhang}
\email{zhangjun@ucas.ac.cn}

\affiliation{Theoretical Physics, Blackett Laboratory, Imperial College London, SW7 2AZ London, United Kingdom}
\affiliation{International Centre for Theoretical Physics Asia-Pacific, Beijing, 100190, China}

\begin{abstract} We consider the effective field theory of gravity around black holes, and show that the coefficients of the dimension-8 operators are tightly constrained by causality considerations. Those constraints are consistent with -- but tighter than -- previously derived causality and positivity bounds and imply that the effects of one of the dimension-8 operators by itself cannot be observable while remaining consistent with causality. We then establish in which regime one can expect the generic dimension-8 and lower order operators to be potentially observable while preserving causality, providing a theoretical prior for future observations. We highlight the importance of ``infrared causality" and show that the requirement of ``asymptotic causality" or net (sub)luminality would fail to properly diagnose violations of causality.
\end{abstract}

\maketitle

\textbf{Introduction.---}
General relativity (GR) should be thought as the leading order term in an effective field theory (EFT) that includes an infinite number of higher-dim operators~\cite{Donoghue:1994dn,Donoghue:1995cz,Burgess:2003jk,Donoghue:2012zc,Weinberg:2016kyd}.
If we are interested in gravity below some energy scale $\Lambda$, we may integrate out all particles with masses above that scale.
Assuming a tree level weakly coupled completion, such as a string theory, the effective action is
\ba
\label{eq:EFT}
{\cal L}_{\rm EFT} = \frac{\mpl^2}{2} \left( R  + \frac{{\cal L}_{\rm D4}}{\Lambda^2} + \frac{{\cal L}_{\rm D6}}{\cut^4} + \frac{{\cal L}_{\rm D8}}{\cut^6} + \cdots
 \right)\,,
\ea
where ${\cal L}_{{\rm D}n}$ denotes a linear combination of all possible dim-$n$ operators built out of the Riemann (or Weyl) curvature and its covariant derivatives, see App.~\ref{app:EFT}.
The higher-dim operators capture the effects of the heavy fields that have been integrated out at tree-level, i.e. particles of spin $\ge2$.
One would expect the scale $\Lambda$ to be the mass of the lightest higher spin state ($s\ge 2$).
Motivated by the recent detections of gravitational waves (GWs), there has been a surge of interest in establishing whether
these operators could be probed assuming a very low $\Lambda$.
Such operators could indicate the presence of new physics beyond the standard model and potentially connect us with the dark sector. A formalism for probing those operators with inspiraling GWs was proposed in Ref.~\cite{Endlich:2017tqa}. Finite size effects of black holes (BHs) have also been investigated in the presence of dim-8 operators \cite{Cardoso:2018ptl} and dim-6 operators \cite{deRham:2020ejn}  \footnote{Motivated by arguments presented in Ref.~\cite{Camanho:2014apa}, when arising from a weakly coupled tree-level completion, the scale of the dim-6 operators is related to the mass of a tower of higher-spin states, which to date has not been observed. Such arguments have motivated focusing instead on dim-8 operators. However it should be pointed out that the precise same arguments equally apply to dim-8 operators and as an EFT,  dim-8 operators seldom appear without the emergence of dim-6 operators at the same scale, unless supersymmetry were preserved.}. Interestingly, LIGO and Virgo constraints on the dim-8 operators were explored in Ref.~\cite{Sennett:2019bpc}. For related works see Refs.~\cite{Brandhuber:2019qpg,Emond:2019crr,AccettulliHuber:2020dal,Cano:2021myl}.

While we may be on the edge of constraining gravitational EFTs using GW observations,  theoretical considerations also have significant impact. For instance, requiring the low-energy EFTs to be embeddable in a local Wilsonian, unitary, Lorentz invariant and causal high energy completion like string theory imposes a set of positivity constraints on these EFTs~\cite{Bellazzini:2015cra,Bern:2021ppb}. In parallel, it is well known that in gravitational EFTs, the sound speed can appear to be superluminal \cite{Drummond:1979pp,Lafrance:1994in,Shore:2002gw,Hollowood:2007kt,Hollowood:2007ku,Hollowood:2008kq,Benakli:2015qlh,
Goon:2016une,deRham:2020ejn,AccettulliHuber:2020oou,Edelstein:2021jyu}, and
by demanding the local group velocity of GWs to be (sub)luminal, it was shown in Ref.~\cite{Gruzinov:2006ie} that the coefficients of the dim-8 operators ought to be sign definite.
In this Letter, we shall complement the state of the art by further investigating the constraints set by causality.
Our requirements for preserving causality are similar to the ones indicated in Refs.~\cite{Hollowood:2015elj,deRham:2020zyh,Chen:2021bvg}  but differ from the notion of ``asymptotic causality" or net (sub)luminality which is sometimes postulated in the literature. As we shall see, the ``asymptotic causality" condition, while necessary is not sufficient for preserving causality and fails to identify situations which are known to be in tension with causality as inferred for instance from positivity bounds.\\

\textbf{GWs in dim-8 EFT.---}
Motivated by the findings of Ref.~\cite{Sennett:2019bpc} we shall start with the following dim-8 operator,
\ba\label{LD81}
S^{(1)}_{\rm{D8}} = \int \d^4 x \sqrt{-g} \frac{\mpl^2}{2} \left[ R + \frac{c_1}{\Lambda^6} \left(R_{abcd}R^{abcd}\right)^2 \right]\,,
\ea
with $c_1=\pm 1$.
Considering a BH of mass $M$, the metric slightly deviates from the Schwarzschild one with a magnitude proportional to the dimensionless parameter $\epsilon=(GM\Lambda)^{-6}$, where $G=1/(8\pi \mpl^2)$ is Newton's constant, see App.~\ref{perturbations} for details.

GWs can be decomposed into odd and even parity metric perturbations $h^{\pm}_{\mu\nu}$ propagating independently on the Schwarzschild-like background. Expressed in spherical harmonics with multipole $\ell$, the radial dependence of each mode can be captured by the master variables $\Psi_{\omega \ell}^{\pm}(r)$, where $\omega$ denotes the frequency. Including  the dim-8 operators  perturbatively, the master variable satisfies  the modified Regge-Wheeler-Zerilli equation \cite{Cardoso:2018ptl}, \footnote{Owing to the spherical symmetry of the background, there is no dependence on the second spherical harmonic quantum number $m$.}
\ba\label{master}
\frac{{\rm d}^2\Psi^\pm_{\omega\ell}}{{\rm d}r_*^2} = -\left[ \omega^2 -V^{\pm}_{\rm GR}(r;\,\ell) - c_1 \epsilon\,  V^\pm (r;\,\ell,\,\omega) \right]\Psi^\pm_{\omega \ell}\,,\quad \
\ea
where $r_*$ is the tortoise coordinate, $V^{\pm}_{\rm GR}$ are the GR potentials  and $V^{\pm}$ the leading-order EFT correction (see App.~\ref{perturbations} for the technical details). Parity ensures that both modes decouple and we shall omit the $\pm$ indices unless relevant.

For the EFT to remain valid when scattering GWs on a BH, the Riemann curvature ought to be small as compared to the cutoff at the impact parameter $r_b = (\ell+1/2)/\omega$, meaning $r_b \Lambda\gg 1$ and $GM/r_b^3\ll \Lambda^2$. Moreover, we also require that the description of the GWs is under control as discussed in Refs.~\cite{deRham:2020zyh,Chen:2021bvg} (see App.~\ref{subsec:regimeValidity}), meaning that
their asymptotic energy $\omega$ should be bounded by
\ba\label{EFTvalidity}
\omega \ll \cut^2 r_b\,.
\ea
With this in mind, the background is then automatically under control if $\epsilon \lesssim 1$ and $r_b>GM$.\\

\begin{figure}[tp]
\centering
\includegraphics[height=0.23\textwidth]{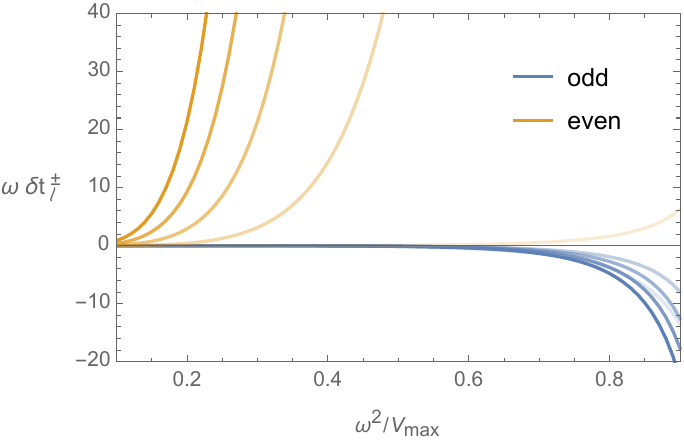}
\caption{EFT corrections on the scattering time delay of the odd (blue) and even (orange) modes in the dim-8 EFT~\eqref{LD81}. From light to dark, the curves show $\omega\, \delta t_\ell$ with $\ell = 2,\, 22,\,42,\,62$ and $82$.
 The EFT contribution to the time delay is given by $\delta T_\ell = c_1 \epsilon \, \delta t_\ell$, so the odd modes enjoy a time advance when $c_1=1$, and the even ones when  $c_1=-1$.
}\label{fig:dtc1}
\end{figure}

\textbf{Scattering phase shift and time-delay.---}
When considering the scattering of GWs on a Schwarzschild-like BH in model~\eqref{LD81}, the EFT corrections manifest themselves in the scattering phase shift and time delay, which can be inferred from solving Eq.~\eqref{master} in the WKB approximation.
For practical reasons, we shall focus on GWs with $\omega^2 <\max(|V_{\rm GR}|)$, in which case the desired WKB solution is the one that decays exponentially at the horizon (tortoise coordinate $r_*\to -\infty$). At infinity, the corresponding solution asymptotes to \cite{deRham:2020zyh},
\ba\label{WKB}
\Psi_{\ell} \propto e^{2i\delta_\ell}e^{i\omega r_*}- (-1)^\ell e^{-i\omega r_*}\,,
\ea
with the phase shift
\ba\label{phase}
\delta_\ell & =&  \int^{\infty}_{r_*^T} {\rm d} r_* \left(\sqrt{\omega^2-V_{\rm GR}- c_1 \epsilon\, V}-\omega \right) \nonumber \\
&& - \omega r_*^T + \frac{\pi}{2} \left(\ell + \frac12\right),
\ea
where $r_*^{T}$ is the turning point defined by $\omega^2 - V_{\rm GR} - c_1 \epsilon\, V = 0$. The scattering time delay  is then given in terms of the phase shift by $T_{\ell} = 2 \pd \delta_\ell (\omega)/\pd \omega$.
As compared to the GR answer $T_{\ell}^{\rm GR}$, the total time delay $T_{\ell}$ acquires an additional EFT contribution $\delta T_\ell$,
\ba
\label{eq:Ttot}
T_{\ell} = T_{\ell}^{\rm GR} + \delta T_\ell + {\cal O}(\epsilon^2)\,.
\ea
Writing $\delta T_\ell = c_1 \epsilon \, \delta t_\ell$, Fig.~\ref{fig:dtc1} shows $\delta t_{\ell}^\pm$ as a function of $\omega$ for various values of $\ell$ (see App.~\ref{app:timedelay}). Interestingly, $\delta t_{\ell}^-$ and $\delta t_{\ell}^+$ always have opposite sign, so that there is always a time advance for one of the GW polarizations for any choice of $c_1=\pm1$.\\

\begin{figure}[tp]
\centering
\includegraphics[height=0.22\textwidth]{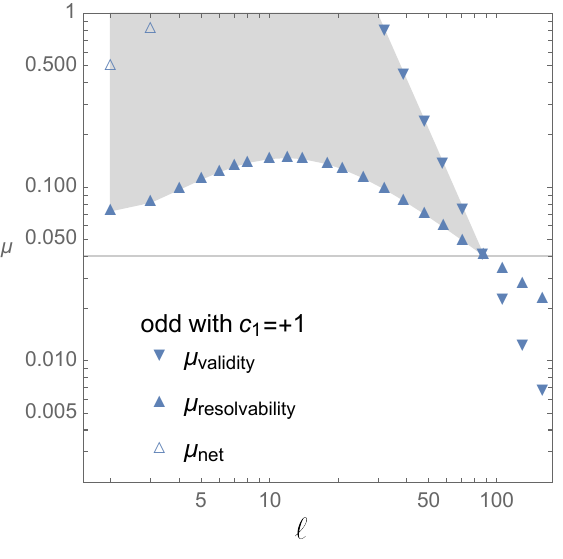} \,
\includegraphics[height=0.22\textwidth]{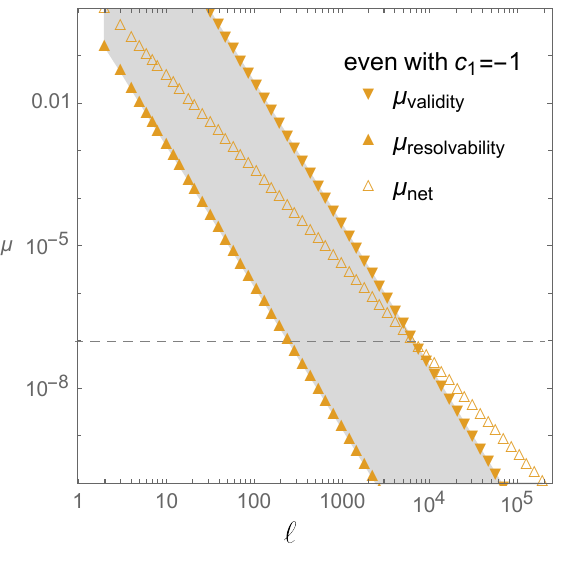}
\caption{Parameter space (shaded grey) of causal violating time advance in the dim-8 EFT~\eqref{LD81}.
$\epsilon_{\rm validity}$ and $\epsilon_{\rm resolvability}$ are the upper and lower bound of $\epsilon$ in condition~\eqref{con}, and $\epsilon_{\rm net}$ is the lower bound in condition~\eqref{con2}.
}\label{fig:regc1}
\end{figure}

\textbf{Infrared Causality.---}
As has been established for QED \cite{Hollowood:2015elj} and for other gravitational theories \cite{deRham:2020zyh}, a time advance compared to GR, i.e. $\delta T_\ell<0$, does not necessarily indicate acausality. It violates causality only if the time advance, calculable {\it within the validity regime of the EFT}, is resolvable. For causality to be respected, the front velocity should be luminal, meaning that the infinite frequency limit of the phase velocity should be luminal as dictated by the geometry seen by those high-frequency modes. As unitarity and analyticity (derived from causality) dictate that the phase velocity cannot decrease with frequency, this implies that low-frequency modes should necessarily be subluminal with respect to the local background geometry. Indeed, the equivalence principle implies that the high-frequency modes can only be sensitive to the local inertial frame, so causality is fixed by the background geometry seen by the high-frequency modes. At the level of a low-energy EFT, causality therefore demands that low-energy modes be (sub)luminal as compared to the background geometry, which in terms of observables requires that any support outside the light cone determined by the geometry be unresolvable, see  \cite{Chen:2021bvg} for more details.
In other words, the statement of `infrared causality' is violated if
\ba\label{resolvable}
-\delta T_\ell \gtrsim 1/\omega \quad \text{(infrared acausality)}.
\ea
Note that infrared acausality necessarily implies the absence of a standard and causal high energy completion, however respecting infrared causality does not necessarily guarantee the presence of a consistent UV embedding, it only is a necessary condition. 
Translating this back into the parameters of the model~\eqref{LD81}, we infer that a wave with frequency $\omega$  and multipole $\ell$ scattered about a BH of mass $M$  in the EFT~\eqref{LD81} violates causality whenever
\ba\label{con}
 \frac{1}{-c_1 \omega\, \delta t_\ell}  \lesssim \epsilon \ll \left(\frac{\ell+1/2}{\omega^2 G^2M^2}\right)^3\,.
\ea
Introducing the parameter $\gamma$ defined as $\omega^2 = \gamma\, V_{\rm max}$, with $V_{\rm max}$ being the maximum of $V_{\rm GR}\sim \ell^2(GM)^{-2}$, the condition~\eqref{con} only depends on the BH mass via $\mu$. Naturally, the effect of the dim-8 operator increases with $\omega$ as illustrated in Fig.~\ref{fig:dtc1}, however $\omega$ should be smaller than $V_{\rm max}$ for the phase shift to be well approximated by \eqref{phase}, \footnote{Violating $\omega^2< V_{\rm max}$ does not indicate a failure of the EFT in any way, it simply indicates that the WKB boundary conditions used to derive \eqref{phase} are no longer valid. In principle we could accommodate for waves with $\omega^2\ge V_{\rm max}$, but such a situation is unlikely to have resolvable time advance within the validity regime of the EFT (see App.~\ref{app:radial}) and is not relevant to our argument.}. For these reasons, we consider $\gamma= 0.9$, so that $\omega \propto \ell$ at large $\ell$, and the impact parameter is constant. We compute the condition~\eqref{con} numerically and present the results in Fig.~\ref{fig:regc1} which indicate that the EFT~\eqref{LD81} violates causality in the odd sector if $c_1 =+1$ and $\epsilon \gtrsim 0.04$.\\

Crucially we see that if $c_1=-1$, the even sector always violates the notion of infrared causality. In that case, the even modes lead to a time advance with $\delta t_\ell^+$ increasing quadratically with the multipole $\ell$, while $\omega$ increases linearly as $\ell \rightarrow \infty$. Therefore, both sides of the inequality~\eqref{con} decrease as $\ell^{-3}$ at large $\ell$. Explicit calculation shows that the left hand side of the inequality~\eqref{con} is smaller than its right hand side (cf. Fig.~\ref{fig:regc1}).
This implies that no matter how small $\epsilon$ is, for sufficiently large $\ell$ the time advance will always be resolvable when $c_1=-1$, hence violates causality,
\footnote{Noting that $V_{\rm GR}^+ \sim \ell^2$ while $V^+ \sim \ell^4$, one may worry that the EFT corrections on the Regge-Wheeler-Zerilli equations might no longer remain perturbative at large $\ell$. However this is not necessary to be case for having causal violating time advance. The time advance is resolvable as long as $-\delta T_\ell \sim \epsilon \ell^2$ is larger than $1/\omega \sim 1/ \ell$, in which case the fractional EFT corrections $V^{+}/V^{+}_{\rm GR} \sim \epsilon \ell^2>1/\ell$ can remain small as $\ell \rightarrow \infty$.}. \\

The main implication of our findings is that the EFT defined  in~\eqref{LD81} can  only ever be causal if $c_1=1$ and if $ (GM\Lambda)^{-6} \lesssim 0.04$ for any BH. Given that the smallest known  BH has $3 M_\odot$ \cite{Thompson:2018ycv,  LIGOScientific:2020zkf, Jayasinghe:2021uqb}, the causality constraint translates into a lower bound on the cutoff scale enforcing $\cut \gtrsim 7 \times 10^{-11} {\rm eV}$.
Within the current state of the art, the EFT~\eqref{LD81} with a cutoff of order $\cut \sim 10^{-13} {\rm eV}$ was shown to lead to observable effects \cite{Sennett:2019bpc}. While such a low cutoff can lead to a potentially interesting phenomenology, it also comes hand in hand with violations of causality. We can push those bounds further by considering  BHs with arbitrarily small mass, which would lead to a constraint on $c_1/\Lambda^6$ to be arbitrarily small. In particular, BHs with radii
as small as the fundamental scale of quantum gravity $M_{\rm Fund}$ would force the scale $\Lambda$ to be associated to that scale $\Lambda \sim M_{\rm Fund}$ in the case where no other dim-6 and -8 operators are considered. \\

Our conclusion on the sign of the coefficient of the dim-8 operator is entirely consistent with expectations on the low-energy operators derived in type II string theory \cite{GROSS19861} after compactification \cite{METSAEV198752}. It also has been shown that the sign of $c_1$ can be fixed by demanding the local group velocity of high-frequency GWs to be (sub)luminal~\cite{Gruzinov:2006ie}. The dim-8 operator $c_1\(R_{abcd}R^{abcd}\)^2$ is, in spirit, the gravitational analog of the $c\, (\partial \phi)^4$ operator that enters generic Goldstone EFTs. In that case the existence of a standard Wilsonian completion, manifests itself via positivity bounds, which have been shown to be directly linked with the sign of the coefficient $c$ \cite{Pham:1985cr,Ananthanarayan:1994hf,Adams:2006sv}. Within the low-energy EFT, the sign of coefficient $c$ is also directly linked to resolvability of time advances and hence to causality \cite{deRham:2020zyh}. Applying similar types of positivity bounds to gravitational EFTs have been shown to impose $c_1>0$ \cite{Bellazzini:2015cra,Bern:2021ppb}.
Not only are our conclusions fully consistent with those results, they also allow us to derive a lower bound of the cutoff of the EFT considered in~\eqref{LD81} when $c_1=1$.
\\

\textbf{Asymptotic~Causality.---}
The time delay $T_{\ell}^{\rm GR}$ introduced in Eq.~\eqref{eq:Ttot}, is the one perceived by the freely propagating modes following null geodesics on that background. In this sense, $T_{\ell}^{\rm GR}$  represents what the high-energy modes (the modes with energy well above $\Lambda$ and well below $\mpl$) are subject to on that very background. See, for instance, Ref.~\cite{Hollowood:2015elj} for an analog discussion in the case of QED.

On the other hand, $\delta T_{\ell}$ represents the additional time delay of the low-energy modes on that same background, arising from interactions with the heavy fields (whose effects are precisely encapsulated by the inclusions of  the higher dimension operators). Since causality demands that low-energy modes do not travel outside the light cone set by the high-energy modes, what matters in setting causality is the sign of $\delta T_\ell$ and not the net $T_{\ell}$. Only a time advance in addition to the GR contribution, i.e., a negative $\delta T_\ell$, would signal that the retarded propagator has support outside the light cone set by the high-energy modes \cite{Shore:1995fz,Shore:2000bs,Hollowood:2009qz,Hollowood:2015elj,deRham:2019ctd,deRham:2020zyh}.

Positivity of the net $T_\ell$, is referred to as asymptotic causality, and violating it requires
\ba
-T_\ell\gtrsim 1/\omega\,, \quad \text{(asymptotic acausality)}\,,
\ea
leading to a different lower bound in the inequality~\eqref{con},
\ba\label{con2}
\frac{(T^{\rm GR}_{\ell}+\omega^{-1})}{-c_1 \delta t_\ell} < \epsilon \ll \left(\frac{\ell+1/2}{\omega^2 G^2 M^2}\right)^3\, .
\ea
From Eq.~\eqref{phase}, we see that $T^{\rm GR}_\ell$ approaches to a constant at large $\ell$, while $\delta t_\ell^{-} \sim \ell^0$ and $\delta t_\ell^{+} \sim \ell^2$, whereas the upper bound in the inequality~\eqref{con2} still scales as $\ell^{-3}$. Therefore, irrespective of the parameters of the EFT, sufficiently high multipoles that remain within the regime of validity of the EFT always enjoy a positive net time delay, as depicted in Fig.~\ref{fig:regc1}. Comparing to the criteria used previously, the statement of asymptotic causality while necessary is not sufficient and by itself would always leads to much weaker constraints on the EFT. For the EFT considered in \eqref{LD81}, the statement of asymptotic causality would allow for a negative $c_1=-1$ so long as $\epsilon <  10^{-7}$, i.e., so long as $\Lambda>15/ GM$. Stated differently, around a BH of mass $M=3 M_\odot$, the net time delay remains positive for all polarizations even if $c_1=-1$ and $\Lambda$ taken to be as low as $10^{-37}\mpl$. Yet we know from positivity bounds that such a situation would be in direct tension with causal and unitary requirements. This illustrates how the statement of asymptotic causality fails to properly diagnose violations of causality. These considerations further show how insisting instead on an unresolvability of the EFT time advance is precisely what is linked with causality considerations in known situations.\\

\textbf{Causality in the generic EFT of gravity.---}
We now generalize the previous argument to more generic gravitational EFTs, and show that the causal requirement $c_1>0$ in model \eqref{LD81} could have been drawn by focusing on the high multipole limit. To keep the discussion general, we consider the EFT corrections on the potential to scale as $V \sim \ell^n$ at large $\ell$, and extend the definition of $\epsilon$ to $\epsilon = (GM \cut)^{-2m}$, where $n$ and $m$ are integers determined by the leading operators present in the EFT. Again, we write $\omega^2 = \gamma\, V_{\rm max}$. As $\ell \rightarrow \infty$, $V_{\rm max} \rightarrow \ell^2/27 G^2M^2$ and hence $\omega \sim \ell$. Focusing on the scaling in $\ell$, the condition~\eqref{con} reduces to
\ba\label{slcon}
\ell^{-n+1} < \epsilon \ll 27^m \ell^{-m}, \quad {\rm for} \quad {\ell \gg 1},
\ea
where the lower bound is the resolvability condition, and the upper bound ensures the  EFT is under control.\\

From the condition \eqref{slcon} we see that a resolvable time advance at infinitely large $\ell$ can only be trusted when $n \ge m+1$. In this case, for any $\epsilon$ there exists a large enough $\ell$, such that the resulting multipole would necessarily violate causality for a particular sign choice of the higher dimensional operator coefficient. This is exactly the case for the even modes in model~\eqref{LD81}, which have $n=4$ and $m=3$. It explains why the sign of $c_1$ has to be definite.
On the other hand the odd modes  have $n=2$, and causality only imposes an upper bound on $\epsilon$.\\

This argument can be directly applied to other higher-dim operators. In particular, the corrected Regge-Wheeler-Zerilli equations in the presence of dim-6 and dim-8 parity-preserving operators takes a similar form as Eq.~\eqref{master}~\cite{deRham:2020ejn,Cardoso:2018ptl}. Up to field redefinitions, there are two additional dim-8 operators beside the one in Eq.~\eqref{LD81}. While one of them is parity violating and is beyond the scope of this Letter, another operator $c_2 (\tensor{\varepsilon}{^a^b_e_f} R_{abcd}R^{efcd})^2$ only affects the odd modes with $V^{-} \sim \ell^4$. In this case, the odd modes exhibit a time advance when $c_2<0$, and the previous argument indicates that causality demands $c_2>0$, which again is fully consistent with the causality requirements inferred in Ref. \cite{Gruzinov:2006ie}~and with the low-energy EFT arising from type II string theory compactification~\cite{GROSS19861,METSAEV198752}.\\

The constraints on the dim-8 EFT implicitly assume that dim-6 ones are subdominant, however up to field redefinitions, the generic EFT of gravity could also include the dim-6 operator $ b_1 \tensor{R}{_a_b^c^d} \tensor{R}{_c_d^e^f} \tensor{R}{_e_f^a^b}$ (see App.~\ref{app:EFT}). In this case, the EFT corrections are suppressed by $\epsilon = (GM\cut)^{-4}$, with $V \sim \ell^2$ at large $\ell$. Performing the same analysis, we find that odd modes can exhibit a resolvable time advance if $b_1=+1$ while even modes can exhibit a time advance if $b_1=-1$. Consistency with causality requires $\Lambda \gtrsim 7.42\times 10^{-10} {\rm eV} (3M_\odot/M) \simeq 419 \mpl^2/M$ (see App.~\ref{app:timedelay}). The statement of infrared causality (imposed by consistency and causality of the UV completion) thus implies that a low-energy EFT of the form \eqref{eq:EFT} can enjoy a causal tree-level weakly-coupled high-energy completion only if the coefficient of the Riemann$^3$ operator vanishes or the cut-off scale is at least above $~ 10^{-9} {\rm eV}$. This is precisely consistent with known explicit string theory realizations. Indeed for maximally supersymmetric and heterotic string theory that coefficient vanishes while it is positive in bosonic string theory \cite{DAppollonio:2015fly}. Note, however, that this result is now proven to be generic for any consistent tree-level weakly-coupled UV completion, independently of the details of the specific realization. Also, this constraint, although obtained by considering the particular dim-6 operator, is imposed on generic EFT with dim-6 operators. \\

\textbf{Observability and Outlook.---}
With the growing interests in probing gravity with GWs, our study provides a theoretical prior from causality considerations for all constraints on EFTs of gravity. Remarkably, for the EFT~\eqref{LD81}, the regime of parameters which was found to be disfavored by the GW events GW151226 \cite{LIGOScientific:2016sjg} and GW170608 \cite{LIGOScientific:2017vox} in Ref.~\cite{Sennett:2019bpc} could have been ruled out on causality considerations alone, assuming a tree level UV completion. This also implies that the current GW observations are not able to test the model~\eqref{LD81} against GR as causality priors require the cutoff of this EFT to be bounded by at least $\Lambda\gtrsim7\times 10^{-11}$eV (possibly much higher). We emphasize that the lower bound on $\Lambda$ is imposed only for the particular dim-8 model~\eqref{LD81}. General dim-8 EFTs usually involve both $c_1$ and $c_2$ operators, and there will be no lower bound on $\Lambda$ from infrared causality considerations if both $c_1$ and $c_2$ are positive and if $c_2/c_1 \sim {\cal O}(1)$ (see App.~\ref{app:timedelay}), in which case the cutoff of the EFT can be as low (or even lower) as that considered in Ref.~\cite{Sennett:2019bpc} and the dim-8 EFTs could be probed/constrained with the current GW observations without being in conflict with infrared causality. Note that If $c_2=0$, and consider BHs with arbitrarily low mass then causality forces $c_1/\Lambda^6$ to be arbitrarily close to zero.

\begin{figure}[tp]
\centering
\hspace{-0.3cm}\includegraphics[height=0.4\textwidth]{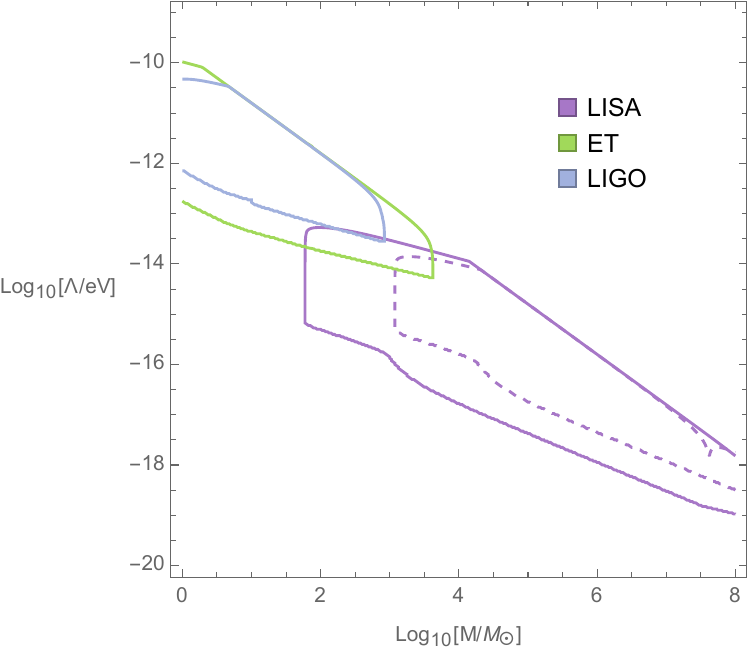}
\caption{Observability of generic dim-6 operators with inspiraling GWs. One could expect to constrain (or observe) dim-6 operators when they enter at a scale $\Lambda$ in the enclosed contours. We consider inspiraling GWs from equal mass binaries BHs with total masses shown in the horizontal axis. For LIGO and Einstein Telescope (ET), we assume the binaries are at $300$Mpc, while for LISA, we consider binaries at $3$Gpc (solid line) and $26$Gpc (dashed line).
}\label{fig:obs}
\end{figure}

We also would like to discuss the possibility of probing gravitational EFTs with dim-6 operators. The dim-6 operators typically dominate over the dim-8 ones in the EFT expansion and could contribute to inspiral waveforms at lower post-Newtonian (PN) order. While dim-6 operators could start contributing to the inspiral waveform at 5PN order~\cite{AccettulliHuber:2020dal}, 3 orders lower than the dim-8 operators (see App.~\ref{app:observability}), we can still use the constraints obtained in Ref.~\cite{Sennett:2019bpc} as a conservative estimation of constraints on the dim-6 EFT corrections. This implies that GW events like GW151226 and GW170608 could already probe the EFT of gravity with dim-6 operators for a cutoff $\cut \in [10^{-13}\,,10^{-12}]$eV or even a wider range. Future GW detectors like Einstein Telescope and LISA are expected to measure the PN coefficients with fractional accuracies of 10\% \cite{Mishra:2010tp, Arun:2006yw} or better. Figure~\ref{fig:obs} shows the potential detectability of generic dim-6 operators with future GW detectors.
Specifically, we assume the dim-6 operators are detectable if their corrections on the phase of observed inspiraling GWs, calculated within the EFT validity regime, is greater than ${\cal O}(1)$. For a given GW source, the EFT corrections are proportional to $\Lambda^{-4}$ and accumulate during inspiral. Therefore, the total dephasing could be less than ${\cal O}(1)$, if $\Lambda$ is too large or is so small that there are not enough GW data available within the EFT validity regime, leading to an upper bound and a lower bound on $\Lambda$ for the dim-6 operators to be detectable. For example, it is possible to probe the dim-6 operators of the EFT of gravity at a cutoff in the range $\cut \in [10^{-14},\,10^{-11}]$ eV if we observe binary BHs of $20 M_\odot$ inspiraling at 300Mpc with the Einstein Telescope. That range can then be lowered to  $\cut \in [10^{-17},\,10^{-15}]$eV if we observe two $10^5 M_\odot$ BHs inspiraling at 3Gpc with LISA. However, these ranges have already been ruled out on causality consideration.

\begin{acknowledgments}
We would like to thank Simon Caron-Huot and Scott Melville for very useful discussions. The work of A. J. T. and C. d.R. is supported by STFC Grants No. ST/P000762/1 and No. ST/T000791/1. C. d.R. thanks the Royal Society for support at ICL through a Wolfson Research Merit Award.  C. d.R. and J. Z. are supported by the European Union Horizon 2020 Research Council Grant No. 724659 MassiveCosmo ERC2016COG. C. d.R. is also supported by a Simons Foundation Award ID 555326 under the Simons Foundation Origins of the Universe initiative, Cosmology Beyond Einstein's Theory and by a Simons Investigator Award No. 690508. J.Z. is also supported by scientific research starting grant No. 118900M061 from University of Chinese Academy of Sciences. A. J. T. thanks the Royal Society for support at ICL through a Wolfson Research Merit Award.
\end{acknowledgments}

\bibliography{references}

\appendix

\onecolumngrid

\section{Higher dimensional operators in gravitational EFTs}\label{app:EFT}
The higher-dimensional operators in gravitational EFT have been constructed, for example, in Ref.~\cite{Ruhdorfer:2019qmk}. The dim-4 operators can be written as
\ba\label{eq:l2}
\mathcal{L}_{\rm D4} &=& a_{R^2} R^2 + a_{W^2} W_{\mu\nu\alpha\beta}^2 + a_{{\rm GB}}R^2_{\rm GB},
\ea
where $R^2_{\rm GB} = R_{\mu \nu \alpha\beta}^2 - 4 R_{\mu \nu}^2 + R^2$ is the Gauss--Bonnet term, and $W_{\mu\nu\alpha\beta}$ is the Weyl tensor. In 4 spacetime dimensions, the Gauss--Bonnet term is topological, which allows us to rewrite the dim--4 curvature operator Lagrangian as
\ba
\label{eq:D4}
\mathcal{L}_{\rm D4} = a_1 R^2 + a_2 R_{\mu\nu}R^{\mu\nu} ,
\ea
with
\ba
a_1 = a_{R^2}- \frac{2}{3} c_{W^2}, \quad  a_2 = 2a_{W^2}.
\ea
The dim--6 operators can be written as
\ba
\label{eq:l3}
\mathcal{L}_{\rm D6} &=&b_1 \tensor{R}{_\mu_\nu^\alpha^\beta} \tensor{R}{_\alpha_\beta^\gamma^\sigma} \tensor{R}{_\gamma_\sigma^\mu^\nu}+ b_2 R^{\mu\nu}R_{\mu\alpha\beta\gamma}\tensor{R}{_\nu^\alpha^\beta^\gamma} +b_{3} R R_{\mu\nu\alpha\beta}R^{\mu\nu\alpha\beta} + b_{4}\tensor{R}{_\mu^\alpha_\nu^\beta}\tensor{R}{_\alpha^\gamma_\beta^\sigma} \tensor{R}{_\gamma^\mu_\sigma^\nu} \nonumber \\
&+&b_5 R\Box R + b_6 R_{\mu \nu} \Box R^{\mu \nu} +b_7 R^3 + b_8 R R_{\mu \nu}^2  + b_9 R_{\mu \nu}^3   + b_{10} R^{\mu \nu} R^{\alpha \beta}R_{\mu\nu\alpha\beta} ,
\ea
Note that the notation is slightly different from the one used in Ref.~\cite{deRham:2020ejn}.

Focusing on a Ricci-flat background, the dim-4 operators and most of the dim-6 operators do not contribute to the linearized equation of GWs, except operators with coefficients $b_1$, $b_2$, $b_3$ and $b_4$ \cite{deRham:2020ejn}. Moreover, in 4-dimension spacetime, the operators with $b_1$ and $b_4$ are related. Namely, one of them can be written as a linear combination of another and the other dim-6 operators \cite{Cano:2019ore}. Therefore, it would be sufficient to consider, for example, operators with $b_1$, $b_2$ and $b_3$ when considering linearized GWs propagating on a Ricci-flat background. Explicit calculation shows that operators with $b_2$ and $b_3$ do not contribute to black hole QNMs~\cite{deRham:2020ejn} nor scattering time delay up to numerical errors. Indeed, if we further allow for field redefinitions, then all operators aside from the one governed by $b_1$ are field redefinable (in the absence of matter). We refer to Ref.~\cite{deRham:2019ctd} for a word of caution in performing field redefinitions when it comes to establishing the speed of various species as the notion of speed is not invariant under field redefinitions even though the notion of causality is. In what follows we shall only focus on the $b_1$ operator so as to better draw an analogue to how dim-8 operators have been treated previously.

\section{BH perturbations}\label{perturbations}
GWs can be considered as metric perturbations propagating on a background metric \ba\label{hmunu}
g_{\mu\nu} = \bar{g}_{\mu\nu} + \frac{1}{\mpl}h^{-}_{\mu\nu} + \frac{1}{\mpl}h^{+}_{\mu\nu}\,,
\ea
where $\bar{g}_{\mu\nu}$ is the background metric, and $h^{\pm}_{\mu\nu}$ are the metric perturbations. Here we have decomposed the metric perturbations into their odd- and even-parity modes, labeled by $-$ and $+$ respectively. For the background, we focus on the static and spherically symmetric BH, which is described by the Schwarzschild metric in GR. In the presence of higher-dimension operators, the Schwarzschild solution usually receives corrections, and the background metric, up to the leading EFT corrections, can be written as \cite{Cardoso:2018ptl, deRham:2020ejn}
\ba\label{BGmetric}
{\rm d} s^2 =- \left[f(r) + \epsilon\, \delta f_t(r) \right] {\rm d} t^2
+ \frac{1}{f(r) +  \epsilon\, \delta f_r(r)} {\rm d} r^2 + r^2 \d \Omega^2\, ,
\ea
where $f(r) = 1- 2GM/r$ with $M$ being the ADM mass of the BH, and $\epsilon\, \delta f_{t/r}$ are the leading EFT corrections with $\epsilon$ being a dimensionless parameter counting the order of the EFT corrections. In the presence of the dim-8 operators, the dimensionless parameter $\epsilon$ is given by $\epsilon = \(GM\cut\)^{-6}$. For any meaningful BH $\epsilon$ is always much less than $1$. For instance, $\epsilon \sim 10^{-8}$ for a $30 M_{\odot}$ BH even assuming a cutoff as incredibly low as $\cut \sim 10^{-10} {\rm eV}$. For reference, in Ref.~\cite{Sennett:2019bpc}, the effects of the EFT~\eqref{LD81} on the waveform of GWs emitted during the inspiral of BHs was considered for a cutoff as low as $\cut \sim 10^{-13} {\rm eV}$, and hence $\epsilon > 1$. In this case, the EFT fails to described the geometry near the BH horizon as well as to predicts the finite-size effects of the BH. Nevertheless, it could be valid in describing the inspiral GWs, as then the energy scale can be well below the EFT cutoff scale. In the case of dim-6 operators, the dimensionless parameter $\epsilon$ is given by $\epsilon = \(GM\cut\)^{-4}$.\\

In order to separate the radial dependence from the time and angular dependence, we can decompose the metric perturbations $h_{\mu\nu}^\pm$ into tensor spherical Harmonics in frequency domain,
 \ba\label{eq:habodd}
h_{\mu\nu}^{-}= e^{-i\omega t}\begin{pmatrix}
0 &0  &- {\rm h}_0 \csc \theta \, \pd_\phi & {\rm h}_0 \sin \theta \, \pd_\theta   \\
0 & 0  & - {\rm h}_1 \csc \theta \, \pd_\phi & {\rm h}_1 \sin \theta \, \pd_\theta  \\
- {\rm h}_0 \csc \theta \, \pd_\phi  & - {\rm h}_1 \csc \theta \, \pd_\phi & \frac{1}{2} {\rm h}_2 \csc \theta \,\mathcal{X} &  -\frac{1}{2} {\rm h}_2 \sin \theta \, \mathcal{W}\\
 {\rm h}_0 \sin \theta \, \pd_\theta &  {\rm h}_1 \sin \theta \, \pd_\theta &  -\frac{1}{2} {\rm h}_2 \sin \theta \,\mathcal{W} &  -\frac{1}{2} {\rm h}_2 \sin \theta \, \mathcal{X}
\end{pmatrix} Y_{\ell m} , \label{odd metric perturbation}
\ea
and
\begin{equation}
h_{\mu\nu}^{+} = e^{-i\omega t}\begin{pmatrix}
- H_0 & \hspace{0.2cm}& H_1  & \hspace{0.2cm}& \mathcal{H}_0 \partial_\theta & \hspace{0.2cm}&  \mathcal{H}_0 \partial_\phi \\
H_1 & \hspace{0.2cm}& H_2  & \hspace{0.2cm}& \mathcal{H}_1 \partial_\theta & \hspace{0.2cm}& \mathcal{H}_1 \partial_\phi \\
\mathcal{H}_0 \partial_\theta & \hspace{0.2cm}& \mathcal{H}_1 \partial_\theta & \hspace{0.2cm}&r^2 \K + r^2 \G \pd_\theta \pd_\theta & \hspace{0.2cm}& r^2 \G \left(\pd_\theta\pd_\phi - \cot \theta \pd_\phi \right)\\
\mathcal{H}_0 \partial_\phi & \hspace{0.2cm}& \mathcal{H}_1 \partial_\phi & \hspace{0.2cm}& r^2 \G \left(\pd_\theta\pd_\phi - \cot \theta \pd_\phi \right) & \hspace{0.2cm}& r^2 \sin^2\theta \K+ r^2 \G \left(\pd_\phi \pd_\phi + \sin\theta \cos \theta \pd_\theta \right)
\end{pmatrix} Y_{\ell m} \, , \label{eq:evenhab}
\end{equation}
where $Y_{\ell m} $ are the spherical harmonics, and
\ba
\mathcal{X} &=& 2(\partial _ \theta \partial _ \phi - \cot \theta \partial _ \phi  )\,, \\
\mathcal{W} &=& (\partial _ \theta \partial _ \theta - \cot \theta \partial _ \theta  - \csc ^2 \theta  \partial _ \phi\partial _ \phi  )\,.
\ea
Here $h_0, h_1, h_2, H_0, H_1, H_2, \mathcal{H}_0, \mathcal{H}_1, \mathcal{K}$ and $\G$ are the radial functions that depend only on $r$. Substituting the above ansatz into perturbation equations leads to a set of equations of the radial functions, which can be further combined into two single master equations. In classical GR, the two master equations are the Regge-Wheeler-Zerilli equations,
\ba\label{masterGR}
\frac{{\rm d}^2\Psi^{\pm}_{\omega\ell }}{{\rm d}r_*^2} = -\left[ \omega^2 -V_{\rm GR}^{\pm}(r;\,\ell) \right]\Psi^{\pm}_{\omega \ell }\,,
\ea
with
\ba
&& V_{\rm GR}^{-} = \frac{f}{ r_g^2}\left( \frac{\lambda+2}{x^2} - \frac{3}{x}\right)\, ,\\
&& V_{\rm GR}^{+}=   \frac{f}{ r_g^2}\left[\frac{\lambda ^2 (\lambda +2) x^3+3 \lambda ^2 x^2+9 \lambda  x+9}{x^3 (\lambda  x+3)^2}\right]\,,
\ea
where $r_g = 2GM$, $x= r/r_g$ and $\lambda = \ell(\ell+1)- 2$.

Including higher dimension operators generally results in higher-derivative terms in the field equations. Within the EFT validity regime, these terms ought be treated perturbatively, and hence can be replaced using the lower order field equations \cite{Cardoso:2018ptl,deRham:2020ejn} (also see Ref.~\cite{deRham:2019ctd} for a generic prescription).
On doing so, we can obtained the Regge-Wheeler-Zerilli equations with the EFT corrections. Up to ${\cal O}(\epsilon)$, the corrected Regge-Wheeler-Zerilli equations can be written as \cite{Cardoso:2018ptl},
\ba\label{app:master}
\frac{{\rm d}^2\Psi^{\pm}_{\omega\ell }}{{\rm d}r_*^2} = -\left[ \omega^2 -V_{\rm GR}^{\pm}(r;\,\ell) - \epsilon\,  V^{\pm} (r;\,\ell,\,\omega) \right]\Psi^{\pm}_{\omega \ell }\,,
\ea
where $r_*$ is tortoise coordinate defined by  ${\rm d}r_*={\rm d}r/\sqrt{(f+ \epsilon\, \delta f_t)(f+\epsilon\, \delta f_r)}$, and $V_{\rm GR}^{\pm}$ are the potentials in the non-corrected Regge-Wheeler-Zerilli equations. Comparing to Eq.~\eqref{master}, we have absorbed the coefficients of the higher-dimension operators in $V^{\pm}$ in Eq.~\eqref{app:master}. For dim-6 operators, we have $\epsilon = 1/(GM\cut)^4$ and
\ba\label{app:master2}
V^{-}=&& \frac{9b_1(1-x)}{x^6} \omega^2 + \frac{b_{1}}{16\, r_g^2\,  x^{10}}  \bigg[ 720 (\lambda -4) x^3 -18 (86 \lambda -643) x^2+(833 \lambda -14723) x \nonumber \\
&& +6024 \bigg], \label{eq:V-}\\
V^{+} =&&\frac{9b_1 (1-x)}{x^6} \omega^2 +  \frac{b_1}{16 \, r_g^2\, x^{10} (\lambda  x+3)^3}
\bigg[-360 (\lambda -4) \lambda ^3 x^6+18 \lambda ^2 \left(44 \lambda ^2-491 \lambda +336\right) x^5 \nonumber \\
&&- \lambda  \left(427 \lambda ^3-14185 \lambda ^2+38106 \lambda -6480\right) x^4 -3 \lambda  \left(2249 \lambda ^2-20515 \lambda +18738\right) x^3\nonumber \\
&&-9 \left(3263 \lambda ^2-11041 \lambda +2646\right) x^2-9 (5495 \lambda -5781) x-28080\bigg] \,.\label{eq:V+}
\ea
In the large $\ell$ limit, we find
\ba
V^{-}_{\ell \rightarrow \infty}=&& \frac{9b_1(1-x)}{x^6} \omega^2 + \frac{b_1\ell^2}{16\, r_g^2\,  x^{9}} \left(720 x^2-1548 x+833\right) \, ,\\
V^{+}_{\ell \rightarrow \infty} =&&\frac{9b_1 (1-x)}{x^6} \omega^2 - \frac{b_1\ell^2}{16\, r_g^2\,  x^{9}}  \left(360 x^2-792 x+427\right)  \,,
\ea
where we keep the $\omega^2$ term as in the main text we consider GWs with $\omega \sim \ell$. Note that for dim-6 operators, the EFT corrections $V^{\pm}$ scales with $\ell^2$ in the large $\ell$ limit.

The dim-8 operator corrections on the Regge-Wheeler-Zerilli equation are derived in Ref.~\cite{Cardoso:2018ptl}. Here we focus on the parity-preserving operators, i.e. the ones with coefficients $c_1$ and $c_2$, corrections from which are given by
\ba
 V^{-}=&& \frac{63 c_1 (x-1)}{x^9} \omega^2 - \frac{1}{8 r_g^2x^{13}}\bigg\{ c_1 \bigg[288 (2 \lambda -77) x^3+(74956-1204 \lambda ) x^2\nonumber \\
 &&+(623 \lambda -83780) x+31005\bigg] + c_2 \bigg[ 144 \lambda  (\lambda +2) (x^3-x^2)  \bigg]\bigg\} \\
 ~\nonumber \\
 V^{+}=&& \frac{63 c_1 (x-1)}{x^9} \omega^2 - \frac{c_1}{8 r_g^2\, x^{13} (\lambda  x+3)^3}  \bigg[36 \lambda ^3 \left(\lambda ^2+18 \lambda +176\right) x^6 \nonumber \\
&&-\,4 \lambda ^2 \left(9 \lambda ^3+238 \lambda ^2+4211 \lambda -7560\right) x^5+\lambda  \left(299 \lambda ^3+16012 \lambda ^2-97068 \lambda +36288\right) x^4 \nonumber \\
&&-\,36 \lambda  \left(154 \lambda ^2-2958 \lambda +3825\right) x^3 -36 \left(1105 \lambda ^2-4757 \lambda +675\right) x^2 \nonumber\\
&&+\,(51948-70020 \lambda ) x-27783\bigg].
\ea
Note that the $c_2$ operator does not contribute to $V^+$. At large $\ell$ we find
\ba
 V^{-}_{\ell \rightarrow \infty}=&& \frac{63 c_1 (x-1)}{x^9} \omega^2 -\frac{1}{8 r_g^2x^{12}}\bigg[ c_1 \ell^2 \left(576 x^2-1204 x+623\right)  + c_2 144 \ell^4 \left(x^2-x\right)\bigg] \\
V^{+}_{\ell \rightarrow \infty} =&&\frac{63 c_1 (x-1)}{x^9} \omega^2  - \frac{ 9 }{2 r_g^2\, x^{11} } c_1\ell^4 (x-1) \, ,
\ea
with $\epsilon = 1/(GM\cut)^6$.

\section{Validity of the EFT}\label{subsec:regimeValidity}
We may use the EFT considered in \eqref{LD81} to study the propagation of GWs on a BH background so long as the effects of other higher dimensional operators do not spoil the predictability of this theory.
As discussed in Ref.~\cite{deRham:2020zyh}, for the EFT to remain valid, the higher-dimension operator corrections, including those from operators with even higher-dimensions must remain perturbative. When comparing to the cutoff scale, these operators, constructed by Riemann curvature and its covariant derivatives, should be evaluated not only on the background but also in presence of GWs. Therefore, in addition to the more familiar constraints on the background curvature (e.g. $W^2 \ll \Lambda^4$) we have
constraints on the GWs which can be characterized by scalars constructed by the momentum of the on shell GWs $k$, the Weyl tensor $W$ and its covariant derivative. For example, considering a transverse wave with $k_\mu = (-\omega,\,0,\,0,\, \pm \omega r^{1/2} \sin \theta /\sqrt{1-r_g/r})$, the EFT validity requires \cite{Chen:2021bvg}
\ba\label{EFTvaliditytran}
\omega \ll \cut^2 r \left(\frac{r}{r_g} -1\right)^{1/2}\, ,
\ea
so that ${\rm Tr}[A^n]\ll \cut^{4n}$, where $\tensor{A}{^a_b} \equiv \tensor{W}{^a_c_b_d}k^ck^d$. For a radial travelling wave with $k_a = (-\omega,\, \pm \omega/(1-r_g/r),\,0,\,0)$, we find $A_{ab}\propto k_a k_b$ and hence the condition ${\rm Tr}[A^n]\ll \cut^{4n}$ becomes trivial. In this case, we need to consider, for example,
\ba
(k^{\mu} \nabla_{\mu})^p (W^{abcd} W_{abcd}) \ll \cut^{4+2p} \, ,
\ea
and the most stringent validity condition is imposed when $p \rightarrow \infty$:
\ba\label{EFTvalidity2}
\omega \ll \cut^2 r.
\ea
Although the partial waves scattering on the BH background are not exactly transverse nor radial travelling, the exact analysis was performed in \cite{Chen:2021bvg}, leading to precisely the same result. In this work, we shall thus use Eq.~\eqref{EFTvalidity} as an appropriate criteria for the validity of the EFT. In particular, the validity condition~\eqref{EFTvalidity} should be imposed at the point of closet approach to the BH for a given partial wave, which can be approximated by the impact parameter $r_b = (\ell+1/2)/\omega$. The validity condition~\eqref{EFTvalidity} also applies when we consider scattering in the presence of other higher-dim operators when considering a generic gravitational EFT.\\

\section{Scattering phase shift and time delay}\label{app:timedelay}
The phase shift of GWs when scattering on a BH can be inferred using the WKB approximation which agrees with the more familiar classical time delay calculations. For GWs with $\omega^2 < |V_{\rm GR}|$, the phase shift is given by Eq.~\eqref{WKB}, and hence the time delay is
\ba
T_\ell =  2 \frac{{\rm d} \delta_\ell}{{\rm d}\omega} = 2 \int^{\infty}_{r^T_*} {\rm d} r_* \left(\frac{2\omega - c_1 \epsilon \frac{\pd V}{\pd \omega}}{2\sqrt{\omega^2-V_{\rm GR}-c_1 \epsilon V}}- 1 \right) - 2 r_*^T\,.
\ea
Expanding to the leading order in $\epsilon$ gives the leading EFT corrections on the time delay,
\ba
\delta T_\ell  =  2 \int^{\infty}_{r^T +\delta{r^T}} {\rm d} r \frac{1}{f+ \epsilon \delta f}\left(\frac{2\omega - c_1 \epsilon \frac{\pd V}{\pd \omega}}{2\sqrt{\omega^2 - V_{\rm GR} - c_1 \epsilon V}} \right)  - 2 \int^{\infty}_{r^T} {\rm d} r \frac{1}{f} \left(\frac{\omega}{\sqrt{\omega^2-V_{\rm GR}}} \right)\,,
\ea
where $r_*^T$, $r^T$ and $\delta r^T$ denotes the turning point in tortoise coordinate, in the usual radial coordinate and the leading EFT corrections on the turning point, and $\delta f$ is defined as
\ba
\epsilon\, \delta f \equiv \sqrt{(f+ c_1\epsilon\, \delta f_t)(f+c_1\epsilon\, \delta f_r)} - f + {\cal O}\,(\epsilon^2)\, .
\ea
To avoid divergence at the turning point, it is useful to define
\ba
&&{\cal A} \equiv \frac{\omega}{f\sqrt{\omega^2-V_{\rm GR}}},\\
&&\delta{\cal A} \equiv  {\cal A} \left[\frac{V}{2 (\omega^2-V_{\rm GR})} -\frac{1}{2\omega}  \frac{\pd V}{\pd \omega} - \frac{\delta f}{f}\right]\,,
\ea
and explicit calculation yields \cite{deRham:2020zyh}
\ba \label{Tadvl}
\delta t_\ell = -2 \int^{\infty}_{r^T} {\rm d} r {\cal A} \left(\frac{\delta {\cal A}}{{\cal A}'}\right)' \, .
\ea

As stated in the letter, avoiding a causal-violating time advance in the model~\eqref{LD81} imposes a lower bound on the cutoff $\Lambda$. We emphasize that the lower bound is imposed only on the model~\eqref{LD81} however general dim-8 EFTs usually involve both the $c_1$ and $c_2$ operators as if for instance the case in type II string theory \cite{GROSS19861} after compactification \cite{METSAEV198752}, where $c_1=c_2>0$. In this case, time advance caused by the $c_1$ operator in the odd sector can be compensated by the time delay caused by the $c_2$ operators when $c_1$ and $c_2$ are positive. As a result, if $c_2/c_1 \sim {\cal O}(1)$, and if $c_1$ and $c_2$ are both positive, there will be no causal violating time advance in the odd sector, nor will there be any lower bound on the cutoff scale $\Lambda$ from infrared causality considerations.

\begin{figure}[tp]
\centering
\includegraphics[height=0.26\textwidth]{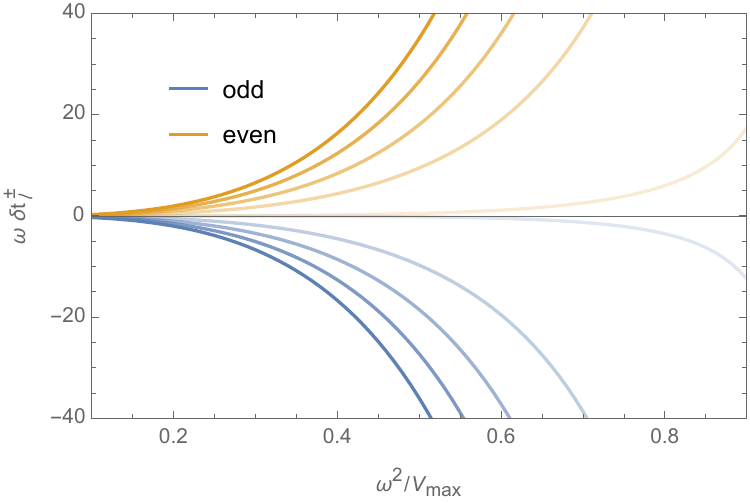} \,
\includegraphics[height=0.26\textwidth]{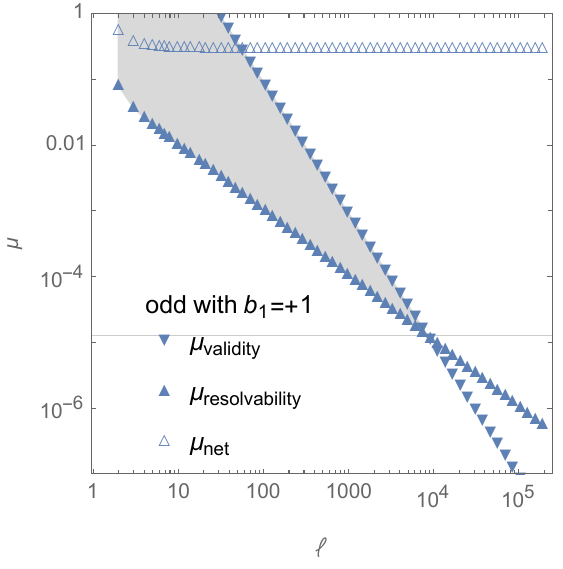} \,
\includegraphics[height=0.26\textwidth]{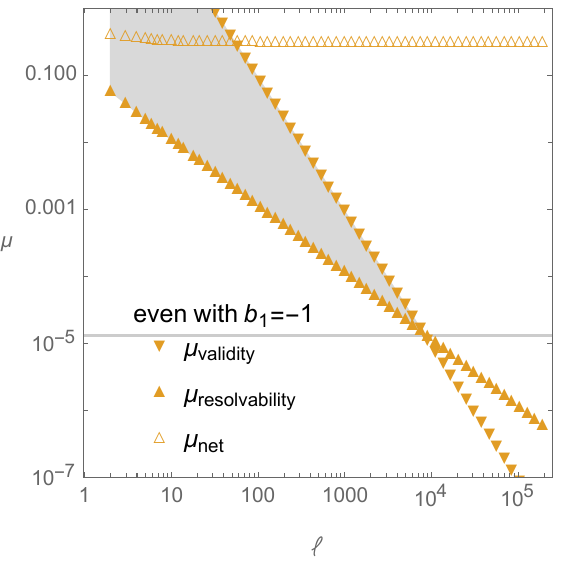}
\caption{The left plot shows the EFT corrections on the scattering time delay of the odd (blue) and even (orange) modes of dim-6 operator with coefficient $b_1$. From light to dark, the curves show $\omega\, \delta t_\ell$ with $\ell = 2,\, 22,\,42,\,62$ and $82$. The EFT contribution to the time delay is given by $\delta T_\ell = b_1 \epsilon \, \delta t_\ell$, so odd modes subject to a positive time delay when $b_1=-1$ and a time advance when $b_1=+1$, while even modes are subject to a positive time delay when $b_1=+1$ and a time advance when $b_1=-1$. The middle and right plots show the parameter space (shaded grey) where dim-6 exhibit a resolvable time-advance that violates the statement of infrared causality. $\epsilon_{\rm validity}$ is the upper bound on $\epsilon$ which ensures validity of the EFT. $\epsilon_{\rm resolvability}$ and $\epsilon_{\rm net}$ are the lower bounds inferred respectively from the statements of infrared and asymptotic causality. We conclude that infrared causality requires $\mu  =(GM\cut)^{-4} < 1.3\times 10^{-5}$.
}\label{fig:regb1}
\end{figure}

This formula also applies for the dim-6 operators. In particular, for the dim-6 operator with coefficient $b_1$, the time delay and the parameter space of causal-violating time advance are shown in Fig.~\ref{fig:regb1}. The dim-6 operators with coefficients $b_2$ and $b_3$ do not contribute to the time delay up to numerical errors.\footnote{This result agrees with that in Ref.~\cite{Melville:2024zjq}. We thank Scott Melville for pointing out a typo of Eq.~\eqref{app:master2} in the previous version.}

~\\

\section{GWs with $\omega \gg V_{\rm max}$}\label{app:radial}
In principle, we could also consider waves with $\omega^2 \gg V_{\rm max}$. Instead of scattering, we now send GWs together with photons from a location near the BH horizon, and compare the time that they travel to asymptotic infinity. In this setup, we expect the time difference is dominated by $\omega$-dependent EFT corrections to the Regge-Wheeler-Zerilli equations. Specifically, we can read off the corrected radial velocity $c_s$ from Eq.~\eqref{master} \cite{Cardoso:2018ptl,deRham:2020ejn},
\ba
c_s^2 = 1 - c_1 \epsilon \left(1-\frac{2GM}{r}\right) \frac{16128 (GM)^8}{r^8}\,.
\ea
The time advance comparing to photons is given by $-\delta T_{\rm rad} = -9c_1 \epsilon GM$ if $c_1 < 0$. The resolvability indicates $-\omega \delta T_{\rm rad} =  -9c_1 \epsilon\, \omega\, GM >1$. On the other hand, the EFT validity requires $GM \omega \ll (G M \cut)^2 $ (cf. Eq.~\eqref{EFTvalidity}), hence we have $ \epsilon\, \omega\, GM \ll \epsilon^{2/3} \ll 1$. Therefore, in this setup, a time advance is unlikely to be resolvable within the validity regime of the EFT.

\section{Observability of higher dimension operators}\label{app:observability}
The direct detection of GWs has sparked interest in probing higher dimension operators entering the EFT of gravity with GW observations. We consider a standard gravitational EFT parameterized as in Eq.~\eqref{eq:EFT}. In the vacuum, the non-topological dim-4 operators have no effect and the EFT corrections are dominated by the dim-6 operators. One can also remove all dim-4 and most of the dim-6 operators with field redefinitions. In the presence of matter fields, field redefinitions  lead to additional couplings between the matter fields, and the dim-4 operators deserve a careful treatment. This is particularly relevant when treating the binaries as point-like particles as well as considering BHs, because the EFT may break down when approaching to the curvature singularities. Nevertheless, we may expect that if the non-vacuum effects of dim-4 operators are calculable within the validity regime of the EFT, $\epsilon$ must be very small and the EFT corrections may no longer be of observational interest. For simplicity, we shall omit these subtleties, and focus on GWs in the vacuum and in the regime where the EFT corrections are dominated by the dim-6 operators. At leading order, we expect that the EFT corrections are proportional to $\epsilon$.

When probing higher dimension operators with GWs from inspirals, the EFT does not have to be valid all the way down to the BH horizon. The inspiral waveform can be constructed within the EFT, so long as the EFT is valid on the lowest length scale in the system, i.e. the binary separation. In this case, $\epsilon = (\Lambda R)^{-4} (R/GM)^4$ could be larger than $1$, where $R$ is the separation of the binary. By considering only early inspiral GWs, one can consider relatively large $\epsilon$ with the price to pay being the loss of available data. Therefore, a theoretical bound on the detectable $\Lambda$ can be estimated as
\ba
\label{eq:boundRL}
\Lambda > (G M \pi f)^{2/3} (GM)^{-1} \,,
\ea
with $f=f_i$ being the initial frequency when the signal becomes detectable. For even lower $\Lambda$, there will be no GW data available within the regime of validity of the EFT.

The upper bound of the detectable $\Lambda$ is determined by the sensitivity of the GW detectors. The EFT corrections on the PN inspiral waveform have been studied in Refs.~\cite{Endlich:2017tqa, AccettulliHuber:2020dal}. Different from the dim-8 ones, the dim-6 operators start contributing to the inspiral waveform at 5PN order \cite{AccettulliHuber:2020dal}. Note that the PN order counting here, different from Ref.~\cite{Endlich:2017tqa}, is based on the frequency dependence. If we consider the case with $\Lambda R \sim 1$, the 5PN EFT corrections, characterised by $\epsilon\, (GM/R)^5 = (\Lambda R)^{-4} (GM/R)$, will be enhanced by a large $\epsilon$, and is numerically equivalent to 1PN effects. If the 5PN coefficient can be constrained with a fraction error of $\delta \hat{p}$ using the inspiralling GWs, we can expect that GWs can constrain $\Lambda$ up to $\Lambda <1/ (\delta \hat{p}^{1/4} GM)$. More concretely, the dim-6 EFT corrections on the phase of inspiralling GWs are roughly
\ba
\Delta \psi^{\rm SPA}_{\rm D6} \sim \mu \left[ \left(GM\pi f_f\right)^{5/3} - \left(GM\pi f_i\right)^{5/3} \right],
\ea
where $f_i$ is the frequency at which the inspiralling GW signal enter the observation band, and $f_f$ is the frequency as which the bound \eqref{eq:boundRL} gets saturated or the maximum frequency of the inspiralling GWs that is observed, whichever is lower. We expect to make a detection of the EFT corrections if $\Delta \psi^{\rm SPA}_{\rm D6} > 1$, which leads to the observability on $\Lambda$ as shown in Fig.~3. Moreover, when limiting ourselves to $\epsilon \ll 1$, we can calculate the EFT corrections on tidal deformability within the EFT validity regime, and search for them in inspiralling GWs.

The higher-dimension operators may also be probed in BH ringdown, for example by measuring quasi-normal frequencies of the BHs, when limiting to $\epsilon \ll 1$. In this case, the EFT corrections on the quasi-normal frequencies are calculable within the EFT validity regime, and are suppressed by $\epsilon$ as comparing to the GR answers \cite{Cardoso:2018ptl,deRham:2020ejn,Cano:2021myl}. The fractional error on BH quasi-normal frequencies measured by the LIGO-Virgo observations is about ${\cal O} (1)$ \cite{LIGOScientific:2020tif}, indicating that the current GW detection cannot probe the higher-dimensional operators with BH ringdown. Future GW detectors are expected to measure QNM with higher accuracy. If the fraction error can be reduced to $\delta\hat{f} \ll 1$, it could allow us to probe the EFT with cutsoff lower than $1/(\delta \hat{f}^{1/4}GM)$, where $M$ is the mass of the observed BH.


\end{document}